# New Candidates for Furin Inhibition as Probable Treat for COVID-19: Docking Output

**Mohammad Reza Dayer**

**Department of Biology, Faculty of Science, Shahid Chamran University of Ahvaz, Ahvaz, Iran**

**Corresponding author:**

**Mohammad Reza Dayer, Department of Biology, Faculty of Sciences, Shahid Chamran University of Ahvaz, Ahvaz, Iran. Tel/Fax: +98-6133331045, E-mail: mrdayer@scu.ac.ir**

**Abstract:**

Furin is a serine protease that takes part in the processing and activation of the host cell pre-proteins. The enzyme also plays an important role in the activation of several viruses like the newly emerging SARS-CoV-2 virus that causes COVID-19 disease with a high rate of virulence and mortality. Unlike viral enzymes, furin owns a constant sequence and active site characteristics and seems to be a better target for drug design for COVID-19 treatment. Considering furin active site as receptor and some approved drugs from different classes including antiviral, antibiotics, and anti protozoa/anti parasites with suspected beneficial effects on COVID-19, as ligands we have carried out docking experiments in HEX software to pickup those capable to bind furin active site with high affinity and suggest them as probable candidates for clinical trials assessments. Our docking experiments show that saquinavir, nelfinavir, and atazanavir with cumulative inhibitory effects of 2.52, 2.16, and 2.13 respectively seem to be the best candidates for furin inhibition even in severe cases of COVID-19 as adjuvant therapy, while clarithromycin, niclosamide, and erythromycin with cumulative inhibitory indices of 1.97, 1.90, and 1.84 respectively with lower side effects than antiviral drugs could be suggested as prophylaxes for the first stage of COVID-19 as a promising treat.

**Keywords:** Furin, COVID-19, Clarithromycin, Erythromycin, Saquinavir, Nelfinavir

**Introduction:**

Furin, EC 3.4.21.75, is a 794 residues serine endoprotease that is encoded by the *FURIN* gene. The enzyme belongs to the subtilisin-like proprotein convertase (PCs) family that catalyzes the hydrolysis of protein substrates at paired basic residues (Arg-X-(Arg/Lys) -Arg, where X can be any amino acid). The enzyme cuts sections from some inactive or precursor proteins and converts them to their active forms. Furin is expressed in all tissues ubiquitously and is mostly found in the trans-Golgi network [1-3].

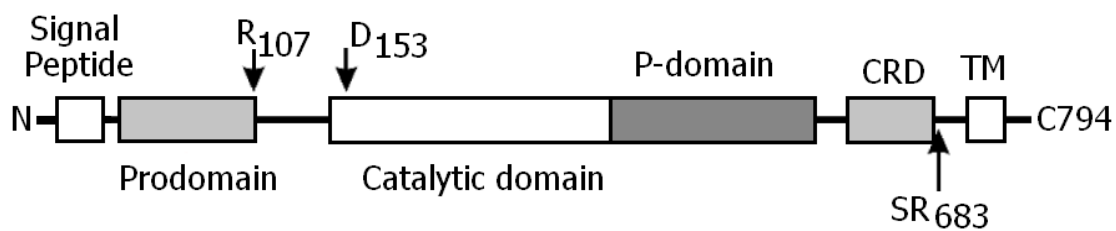

**Figure-1:** Primary structure of furin enzyme, from N-terminal includes, signal peptide, prodomain, catalytic domain, p-domain, CRD (cysteine rich domain) and TM (transmembrane domain) respectively.

Figure 1 represents the primary structure of the inactive form of furin on synthesis. As it is depicted furin comprises a short signal peptide at the N-terminal followed by a prodomain ending at a cleavage site of R107 upon activation. The prodomain helps correct the folding of the next catalytic domain initially. This domain is also known as inhibitory propeptide since it inhibits the proteolytic activity of furin until its removal. This domain is cleaved off by autoproteolytic of activated forms of furin or by other subtilisin-like proteases. The next domain is called the catalytic domain which contains an essential triad of Asp, His, and Ser that take part in proteolytic activity. P-domain is the following domain playing role in regulating enzyme activity at different pH and calcium ions concentrations [4-5]. Cysteine-rich domain (CRD) is the next functional domain ending with the next cleavage site SR683 residues. At certain conditions, this site is also cleaved off by active furin or by other subtilisin-like enzymes. Under this condition furin leeks to the extracellular compartment with retained activity. The ultimate domain at the C-terminal site of furin is the transmembrane (TM) domain that tight the enzyme to membranes of the endoplasmic reticulum [6-7].

Besides the physiological role, furin is also recruiting to process some pathogens proteins such as envelope proteins of HIV, influenza, and several filoviruses of Ebola and Marburg virus and also spike protein of SARS-CoV-2 and therefore fully activates the pathogens [8-9]. The tertiary structure of furin, Figure 2 (left), shows the p-domain beings at the C-terminal part and the catalytic domain being at the N-terminal portion of the protein. Furin catalytic domain, Figure 2 (right), contains active site cavity lined by negative charge residues, where the enzyme substrates or inhibitors can bind the enzyme and come to contact with the catalytic triad of Asp153, His194, and Ser368 which is important for catalysis. The presence of negatively charged residues in the catalytic site explains the requirement for substrates/inhibitors to carry positive charges for effective binding and also explains the cleavage point of proteins to be at positive residues of Lys and/or Arg [10-11]. As it is indicated in figure 2 there are two or three calcium ions depend on the origin of furin that is not involved in the catalytic process but are essential for enzyme native conformation [12-13].

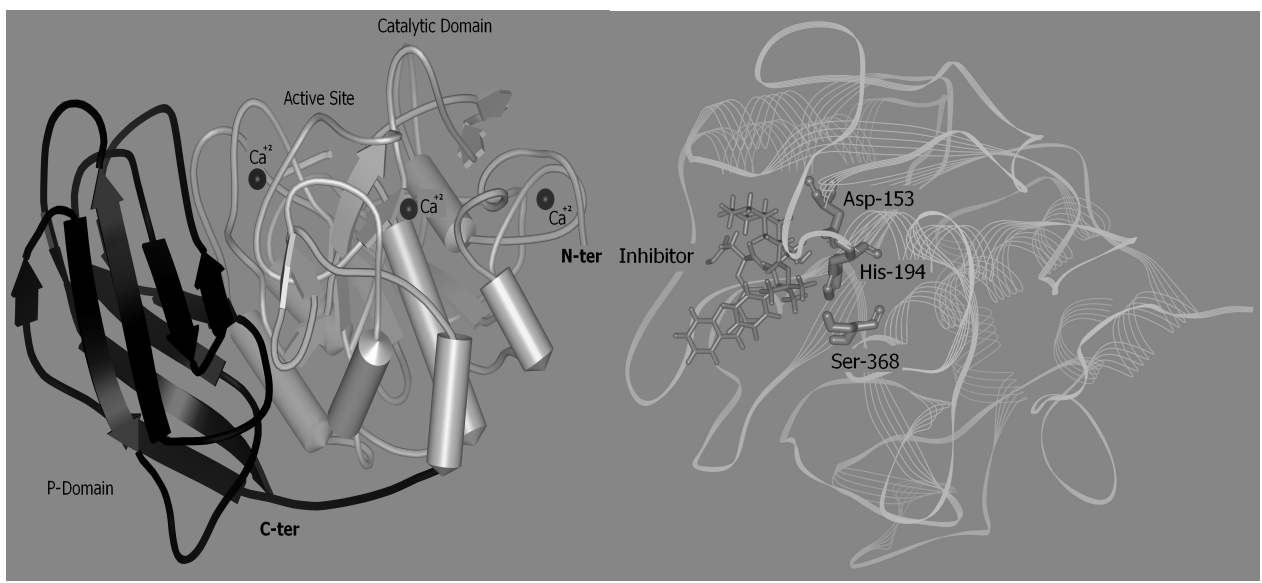

**Figure 2:** Left, furin tertiary structure obtained from crystallography studies at 2 angstrom resolutions, right, furin active site occupied by an inhibitor

It is well documented that furin is up-regulated in several conditions, like diabetes mellitus, cancer, and viral infections that is suspected to play role in *diseases deterioration* and so comprises a potential target for drug development and inhibition [14-15]. COVID-19 among the newly emerging viral disease that is caused by SARS-CoV-2 comprises a serious pandemic worldwide with more than 194 million cases and 4.16 million deaths by Jul 2021. The diseases encourage investigators to search for a vaccine for healthy individuals or drugs for patients. It is shown that extracellular furin plays a critical role in SARS-CoV-2 infectivity [16-17]. The virus uses its surface spike or S protein to binds the host cell receptor called angiotensin-converting enzyme 2 (ACE2). This protein contains N-terminal or S1 domain, which is responsible for receptor binding, and a C-terminal or S2 domain, for host cell fusion and entrance. Furin by cleaving the spike at S1/S2 cleavage site accelerates virus entry and pathogenesis [18-20].

One strategy against viral infections is the application of inhibitors against viral enzymes to inhibit viral amplification. The main obstacle in this way is the drug resistance posed by a high rate of replication errors. Inhibition of furin with no drug resistance seems to be a good approach to prevent viral infections especially in the case of COVID-19 [21-22]. According to this introduction and during this work we tried to search for drugs capable to bind furin active site and inhibition among known anti-viral and approved antibiotics based on their structural similarities through molecular docking calculations with the hope to find and suggest new candidates to fight the SARS-CoV-2 virus.

**Materials and Methods:**

**Furin coordinate structure:**

The coordinate structure of furin with PDB ID of 4OMD was retrieved from the protein data bank (https://www.rcsb.org/). The structure was obtained by the X-ray diffraction methods and refined at the resolution of 2.70 Å. Since the protein structure was refined from dried crystal, so its conformation will be far from its native structure in physiological conditions. Accordingly, we optimized and equilibrate the protein structure via minimization to under 200kJ/mol in pH 7.5, 37 degrees centigrade, and 1atomsphere of pressure using in GROMACS 4.5.5 software (http://www.gromacs.org) and GROMOS force field and steepest descent algorithm. The structure was placed in a rectangular box with a dimension of 6.11×7.48×7.27nm filled with SPCE water [23-24].

**Coordinate structures of inhibitors:**

The chemical structures of candidate inhibitors, including antiviral drugs of amprenavir, atazanavir, baloxavir, darunavir, disoproxil, emtricitabine, indinavir, lamivudine, lopinavir, nelfinavir, nevirapine, oseltamivir, remdesivir, ritonavir, saquinavir, tenofovir, tipranavir, zidovudine; antibiotics of azithromycin, cefaclor, cefazolin, cefdinir, cefditoren, cefixime, cefotaxime, cefpodoxime, cefprozil, ceftizoxime, ceftriaxone, cefuroxime, ciprofloxacin, clarithromycin, doxycycline, erythromycin, fidaxomicin, gemifloxacin, imipenem, moxifloxacin, ofloxacin, sulfamethoxazole, tetracycline; anti protozoa/anti parasite of diminazene, i, and niclosamide, in SDF format were obtained from PubChem database (https://pubchem.ncbi.nlm.nih.gov/) and converted to PDB format in Open Babel software (http://openbabel.org/) and energy minimized in ArgusLab software (http://www.arguslab.com/) [25].

**Furin Active Sites:** Furin's active site was extracted from the PDB structure of 4OMD i.e., the binding site of competitive inhibitor using ArgusLab software (http://www.arguslab.com/) [25].

**Molecular docking experiments:** To carry molecular docking experiments the optimized structure of furin with PDBID:4OMD was used as a receptor and the coordinate structures of inhibitor in PDB formats were used as a ligand for blind docking experiments in Hex 8.0.0 (http://www.loria.fr/~ritchied/hex/) software [26]. The sahpe+electrostatic and macro sampling modes of docking were used for docking and the best 100 poses and their binding energies were saved for further analysis.

**Drugs Partition coefficient:** The logP or partition coefficient for drugs is a known index for hydrophobicity, the more and positive values reflecting the more hydrophobic nature for a chemical and vice versa [27]. The Virtual Computational Chemistry Laboratory (http://www.vcclab.org/) server was used to calculated logP for each drug [28].

**Data Handling and Analysis:** All the numerical data were exported to Excel and SPSS software for analysis. A P-value lower than 0.05 was considered as the significance level.

**Results and Discussion:**

It is well known that interruption of S1/S2 cleavage of S protein of SARS-CoV-2 by host protease and especially furin using inhibitors can prevent virus entry and pathogenesis [28-30]. Accordingly, furin becomes a good target for drug design against viral infections especially for the management of the COVID-19 pandemic threat currently [31-32]. Considering the resolved crystal structure of furin, the in silico molecular docking experiments were conducted to predict the ability of small drugs approved for clinical application molecule as candidates for furin inhibitions and COVID-19 treatment. Furin inhibitors may bind as competitive inhibitors to the furin active site or to the interface cleft between P-domain and catalytic site as non-competitive inhibitors. Since the binding site seems to be a better target to followed precisely in this work our goal was to evaluate the binding capacity of the tested drugs to bind the enzyme active site and their potential was judged based on their binding energy (-ΔG).

To perform docking experiments with furin as a receptor, we have selected the crystal structure of furin with PDB ID of 4OMD from the protein data bank. To assure that the crystal structure matches the wild-type protein and to survey their uniformity we performed pair-wise alignment on emboss sever (www.ebi.ac.uk/Tools/psa/emboss_needle/). As it is depicted in figure 3, 4OMD structure completely has a wild-type structure without any mutations.

```
Wile-Type          6 D--------------------S----------------------CNCDG     12
                     |                    |                    |||||
PDBID:4OMD       151 DDGKTVDGPARLAEEAFFRGVSQGRGGLGSIFVWASGNGGREHDSCNCDG   200

Wile-Type         13 YTNSIYTLSISSATQFGNVPWYSEACSSTLATTYSSGNQNEKQIVTTDLR    62
                     |||||||||||||||||||||||||||||||||||||||||||||||||
PDBID:4OMD       201 YTNSIYTLSISSATQFGNVPWYSEACSSTLATTYSSGNQNEKQIVTTDLR   250

Wile-Type         63 QKCTESHTGTSASAPLAAGIIALTLEANKNLTWRDMQHLVVQTSKPAHLN   112
                     |||||||||||||||||||||||||||||||||||||||||||||||||
PDBID:4OMD       251 QKCTESHTGTSASAPLAAGIIALTLEANKNLTWRDMQHLVVQTSKPAHLN   300

Wile-Type        113 ANDWATNGVGRKVSHSYGYGLLDAGAMVALAQNWTTVAPQRKCIIDILTE   162
                     |||||||||||||||||||||||||||||||||||||||||||||||||
PDBID:4OMD       301 ANDWATNGVGRKVSHSYGYGLLDAGAMVALAQNWTTVAPQRKCIIDILTE   350

Wile-Type        163 PKDIGKRLEVRKTVTACLGEPNHITRLEHAQARLTLSYNRRGDLAIHLVS   212
                     |||||||||||||||||||||||||||||||||||||||||||||||||
PDBID:4OMD       351 PKDIGKRLEVRKTVTACLGEPNHITRLEHAQARLTLSYNRRGDLAIHLVS   400

Wile-Type        213 PMGTRSTLLAARPHDYSADGFNDWAFMTTHSWDEDPSGEWVLEIENTSEA   262
                     |||||||||||||||||||||||||||||||||||||||||||||||||
PDBID:4OMD       401 PMGTRSTLLAARPHDYSADGFNDWAFMTTHSWDEDPSGEWVLEIENTSEA   450

Wile-Type        263 NNYGTLTKFTLVLYGTAPEG-LPVPPESSGCKTLTSSQACVVCEEGFSLH   311
                     ||||||||||||||||| .| ||        :            | | |
PDBID:4OMD       451 NNYGTLTKFTLVLYGTA-SGSL-VP------R-------------G-S-H   477
```

**Figure 3:** sequence alignment result for 4OMD in contrast to its wild type sequence showing no mutation in 4OMD structure (www.ebi.ac.uk/Tools/psa/emboss_needle/)

Table 1 represents the docking results including the percent of binding site occupation and the binding energies in kJ/mol for our tested drugs. It should be mentioned that baloxavir, cefaclor, cefdinir, cefotaxime, cefuroxime, cefpodoxime, ciprofloxacin, emtricitabine, gemifloxacin, imipenem, moxifloxacin, ofloxacin, sulfamethoxazole, and tipranavir could not bind to enzyme active site (0%) and were deleted from further evaluation, however, they may bind to an allosteric site and inhibit the enzyme non-competitively. Table 1 also depicts the logP for drugs which reflects the hydrophobicity or bioavailability of drugs to reach the furin as a target for inhibition.

**Table 1:** Percent of active site occupation, logP, and binding energies (kJ/mol) extracted from our docking experiments

|  | % of occupation | logP | Binding Energy |
|---|---|---|---|
| Remdesivir | 5 | 2.2 | -441.90 |
| Tetracycline | 7 | -0.56 | -362.57 |
| Zidovudine | 8 | -0.1 | -326.77 |
| Darunavir | 10 | 1.89 | -348.13 |
| Lopinavir | 10 | 3.91 | -280.45 |
| Nevirapine | 11 | 1.75 | -383.61 |
| Azithromycin | 11 | 3.03 | -401.92 |
| Lamivudine | 13 | -1.29 | -472.23 |
| Oseltamivir | 13 | 1.3 | -280.76 |
| Ceftizoxime | 14 | 0.4 | -296.96 |
| Cefazolin | 15 | -0.4 | -316.74 |
| Ceftriaxone | 20 | -0.01 | -335.60 |
| Diminazene | 20 | 1.09 | -295.01 |
| Tenofovir | 20 | -1.51 | -245.58 |
| Doxycycline | 23 | -0.72 | -328.83 |
| Cefditoren | 28 | 1.7 | -357.90 |
| Ritonavir | 33 | 4.24 | -359.53 |
| Ivermectin | 37 | 4.04 | -277.68 |
| Amprenavir | 41 | 2.03 | -347.75 |
| Erythromycin | 44 | 2.37 | -382.81 |
| Clarithromycin | 47 | 3.18 | -382.47 |
| Niclosamide | 47 | 4.49 | -272.53 |
| Atazanavir | 51 | 4.08 | -395.35 |
| Indinavir | 56 | 3.26 | -273.01 |
| Nelfinavir | 62 | 4.61 | -288.81 |
| Saquinavir | 86 | 4.04 | -374.82 |

A more hydrophobic drug (higher logP) with higher binding energy in accordance with more percent of binding site occupation will be a more effective inhibitor. In order to make a reasonable comparison we normalized the calculated percent of occupation, logP, and binding energies values and summate them in a cumulative index. Therefore the more cumulative index reveals the more effective drug for furin inhibition.

**Table 2: Normalized values for active site occupation, logP, and binding energy in accordance with extracted cumulative index**

| | % of occupation | logP | Energy | Cumulative |
|---|---|---|---|---|
| Tenofovir | 0.23 | 0.04 | 0.52 | 0.79 |
| Cefixime | 0.06 | 0.26 | 0.67 | 0.99 |
| Zidovudine | 0.09 | 0.21 | 0.69 | 1.00 |
| Tetracycline | 0.08 | 0.16 | 0.77 | 1.00 |
| Cefazolin | 0.17 | 0.18 | 0.67 | 1.02 |
| Cefprozil | 0.03 | 0.34 | 0.65 | 1.02 |
| Ceftizoxime | 0.16 | 0.27 | 0.63 | 1.07 |
| Doxycycline | 0.27 | 0.14 | 0.70 | 1.10 |
| Oseltamivir | 0.15 | 0.38 | 0.59 | 1.13 |
| Ceftriaxone | 0.23 | 0.22 | 0.71 | 1.17 |
| Diminazene | 0.23 | 0.36 | 0.62 | 1.22 |
| Lamivudine | 0.15 | 0.07 | 1.00 | 1.22 |
| Darunavir | 0.12 | 0.46 | 0.74 | 1.31 |
| Nevirapine | 0.13 | 0.44 | 0.81 | 1.38 |
| Lopinavir | 0.12 | 0.71 | 0.59 | 1.42 |
| Remdesivir | 0.06 | 0.50 | 0.94 | 1.49 |
| Cefditoren | 0.33 | 0.43 | 0.76 | 1.52 |
| Azithromycin | 0.12 | 0.59 | 0.85 | 1.57 |
| Amprenavir | 0.48 | 0.47 | 0.74 | 1.69 |
| Ivermectin | 0.43 | 0.72 | 0.59 | 1.74 |
| Erythromycin | 0.51 | 0.52 | 0.81 | 1.84 |
| Indinavir | 0.65 | 0.63 | 0.58 | 1.86 |
| Ritonavir | 0.38 | 0.75 | 0.76 | 1.89 |
| Niclosamide | 0.55 | 0.78 | 0.58 | 1.90 |
| Clarithromycin | 0.55 | 0.62 | 0.81 | 1.97 |
| Nelfinavir | 0.72 | 0.79 | 0.61 | 2.13 |
| Atazanavir | 0.59 | 0.73 | 0.84 | 2.16 |
| Saquinavir | 1.00 | 0.72 | 0.79 | 2.52 |

As it is evident saquinavir, atazanavir, and nelfinavir, obtain a higher score in our series with 2.52, 2.16 and 2.13 cumulative indices respectively. In vitro assays show that nelfinavir exerts more effects on viral infections in contrast to ritonavir and lopinavir, while there is no experimental evidence regarding the clinical application of atazanavir and saquinavir in viral disease [33-35]. Among the rest anti viral drugs used in this work, ritonavir, indinavir, amprenavir, remdesivir, lopinavir, nevirapine, darunavir, lamivudine, oseltamivir, zidovudine and tenofovir with cumulative indices varies between 1.89 to 0.79 retain the next order of effectiveness in furin inhibition. It is important to notice that darunavir as example has least affinity to attacks furin active site while it can inhibits furin by binding to its allosteric site with high enough, this confirms our finding that darunavir only in 10 percent of probability binds furin active site [36]. Unfortunately, there is no in vitro document for their binding potency for the rest of anti viral drugs. Macrolides used in our study including azithromycin, erythromycin and clarithromycin indicate with cumulative indices of 0.99, 1.84 and 1.97 respectively reveal that they especially incase of erythromycin and clarithromycin merit valuable importance for further studies in case of COVID-19 treatment. There increasing reports regarding the importance of macrolides in COVID-19 treatment [37-39]. Niclosamide among anti protozoa/parasite drugs used in this work include diminazene, niclosamide and ivermectin with cumulative index of 1.90 seems to be good candidate for clinical trail, however there are reports implying the positive effects of ivermectin in COVID-19 [40-42]. Table 3 summarize the interactions of drugs with their counterpart residues in furin active site analyzed and extracted on Protein-Ligand Interaction Profiler (PLIP) server (https://plip-tool.biotec.tu-

dresden.de/plip-web/plip/index). This server gives all non-covalent interactions including, hydrogen bonds, salt bridges, hydrophobic interactions, and other non covalent interaction for docked complexes. As it is evident all the selected drugs make hydrophobic and hydrogen bond interactions with catalytic residues of the enzyme and their neighbors residues that make them effective binders for furin inhibition.

Table-3: Interactions between drugs attached to furin active site were analyzed on PLIP server (https://plip-tool.biotec.tu-dresden.de/plip-web/plip/index). H-A (hydrogen-acceptor distance) and H-D (hydrogen-donor distance)

| Inhibitor | Residue | Distance (Angstrom) | Type of interaction |
|---|---|---|---|
| Atazanavir | Leu-227 | 3.42 | Hydrophobic |
| Atazanavir | Glu-257 | 3.86 | Hydrophobic |
| Atazanavir | Pro-256 | $3.45_{(H-A)}$, $4.06_{(D-A)}$ | **Hydrogen Bond** |
| Atazanavir | Asp-258 | $3.06_{(H-A)}$, $3.48_{(D-A)}$ | **Hydrogen Bond** |
| Atazanavir | Asn-295 | $3.04_{(H-A)}$, $3.78_{(D-A)}$ | **Hydrogen Bond** |
| Atazanavir | Ser-368 | $3.28_{(H-A)}$, $4.05_{(D-A)}$ | **Hydrogen Bond** |
| Nelfinavir | Arg-193 | 3.79 | Hydrophobic |
| Nelfinavir | Val-231 | 3.54 | Hydrophobic |
| Nelfinavir | Asp-153 | $3.43_{(H-A)}$, $3.89_{(D-A)}$ | **Hydrogen Bond** |
| Nelfinavir | His-194 | $2.80_{(H-A)}$, $3.24_{(D-A)}$ | **Hydrogen Bond** |
| Clarithromycin | Asp-191 | 3.67 | Hydrophobic |
| Clarithromycin | His-194 | 3.59 | Hydrophobic |
| Clarithromycin | Asp-154 | $2.35_{(H-A)}$, $3.32_{(D-A)}$ | **Hydrogen Bond** |
| Clarithromycin | Asp-258 | $3.47_{(H-A)}$, $3.99_{(D-A)}$ | **Hydrogen Bond** |
| Clarithromycin | Asp-258 | $1.94_{(H-A)}$, $2.88_{(D-A)}$ | **Hydrogen Bond** |
| Clarithromycin | Asp-258 | $2.27_{(H-A)}$, $3.07_{(D-A)}$ | **Hydrogen Bond** |
| Clarithromycin | Asn-295 | $2.15_{(H-A)}$, $3.12_{(D-A)}$ | **Hydrogen Bond** |
| Clarithromycin | Ser-368 | $3.37_{(H-A)}$, $3.87_{(D-A)}$ | **Hydrogen Bond** |
| Erythromycin | Asp-154 | 3.87 | Hydrophobic |
| Erythromycin | Leu-227 | 3.60 | Hydrophobic |
| Erythromycin | Trp-254 | 3.82 | Hydrophobic |
| Erythromycin | Asp-258 | 3.87 | Hydrophobic |
| Erythromycin | Asp-191 | $1.87_{(H-A)}$, $2.75_{(D-A)}$ | **Hydrogen Bond** |
| Erythromycin | Asp-258 | 4.36 | **Salt Bridge** |
| Niclosamide | Asp-154 | 3.92 | Hydrophobic |
| Niclosamide | His-194 | 3.78 | Hydrophobic |
| Niclosamide | Leu-227 | 3.57 | Hydrophobic |
| Niclosamide | Trp-254 | 3.90 | Hydrophobic |
| Niclosamidea2 | Thr-367 | 3.40 | Hydrophobic |
| Niclosamide | Leu-227 | $1.82_{(H-A)}$, $2.80_{(D-A)}$ | **Hydrogen Bond** |
| Saquinavir | Arg-193 | 3.74 | Hydrophobic |
| Saquinavir | Leu-227 | 3.93 | Hydrophobic |
| Saquinavir | Trp-254 | 3.85 | Hydrophobic |
| Saquinavir | Asp-154 | 5.13 | **Salt Bridge** |

**Conclusion:** Considering the maximum allowed dose for drugs and their cumulative indices as well as their biocompatibility and side effects we suggest erythromycin with more than 1gr/day and clarithromycin with 0.4gr/day among macrolides at the first step of disease as prophylaxis over niclosamide (with more than 1gr/day), and antiviral drugs of saquinavir, nelfinavir, indinavir and atazanavir with less than 1gr/day in sever state of the disease with the higher most cumulative indices.

**Acknowledgements**

The author would like to express his thanks to the vice-chancellor of research and technology of the Shahid Chamran University of Ahvaz for providing financial support for this study under Research Grant No: SCU.SB99.477.